\documentclass[twoside,11pt]{article}

\usepackage{jmlr2e}
\usepackage{subfigure}
\usepackage{float}
\usepackage{lineno}
\usepackage[colorlinks=false]{hyperref}%<--inserisce segnalibri e riferimenti indice
\usepackage{natbib}
\usepackage{cleveref}% l
\usepackage{graphicx}

\ShortHeadings{Bayesian dynamic financial networks with time-varying predictors}{}

\begin{document}

\title{Bayesian dynamic financial networks with time-varying predictors}

\author{\name Daniele Durante \email durante@stat.unipd.it\\
      \addr Department of Statistical Sciences\\
       University of Padua\\
       Padua, Italy
\AND
       \name David B. Dunson \email dunson@duke.edu \\
       \addr Department of Statistical Science\\
     Duke University\\
       Durham, NC 27708-0251, USA
}

\editor{}

\maketitle

\begin{abstract}%
We propose a Bayesian nonparametric model including time-varying predictors in dynamic network inference.  The model is applied to infer the dependence structure among financial markets during the global financial crisis, estimating effects of verbal and material cooperation efforts. We interestingly learn contagion effects, with increasing influence of verbal relations during the financial crisis and opposite results during the United States housing bubble. 
\end{abstract}
\begin{keywords}
Co-movement; Edge Covariates; Financial Network; Gaussian Process; Latent Space; Matrix Factorization.
\end{keywords}
\section{Introduction}
The global financial crisis is a hotly debated topic having complex roots and ongoing effects.  The crisis was driven by the coexistence of a complex system of causes covering easy credit conditions, lack of proper regulation and the introduction of new financial instruments. The 2004-2007 United States housing bubble is a key factor behind the subsequent financial instability, tracing back its causes to the unusually low interest rates decision of the Federal Reserve to soften the effects of the $2000$ dot-com bubble \citep{tay}, and the growing demand for financial assets by foreign countries which generated an additional influx of ``saving glut'' \citep{ber} stimulating the proliferation of risky loans, predatory lending, increasing financial complexity \citep{bru} and a wide network of dependencies between financial operators worldwide. Optimistic forecasts on the expansion of the real estate market contributed to the inflation of the bubble, which burst between 2006 and 2007, when subprime borrowers proved unable to repay their mortgage, triggering a vicious cycle with owners holding negative equity motivated to default on their mortgages.
\begin{figure}[t]
\centering
\includegraphics[height=3.7cm, width=10cm]{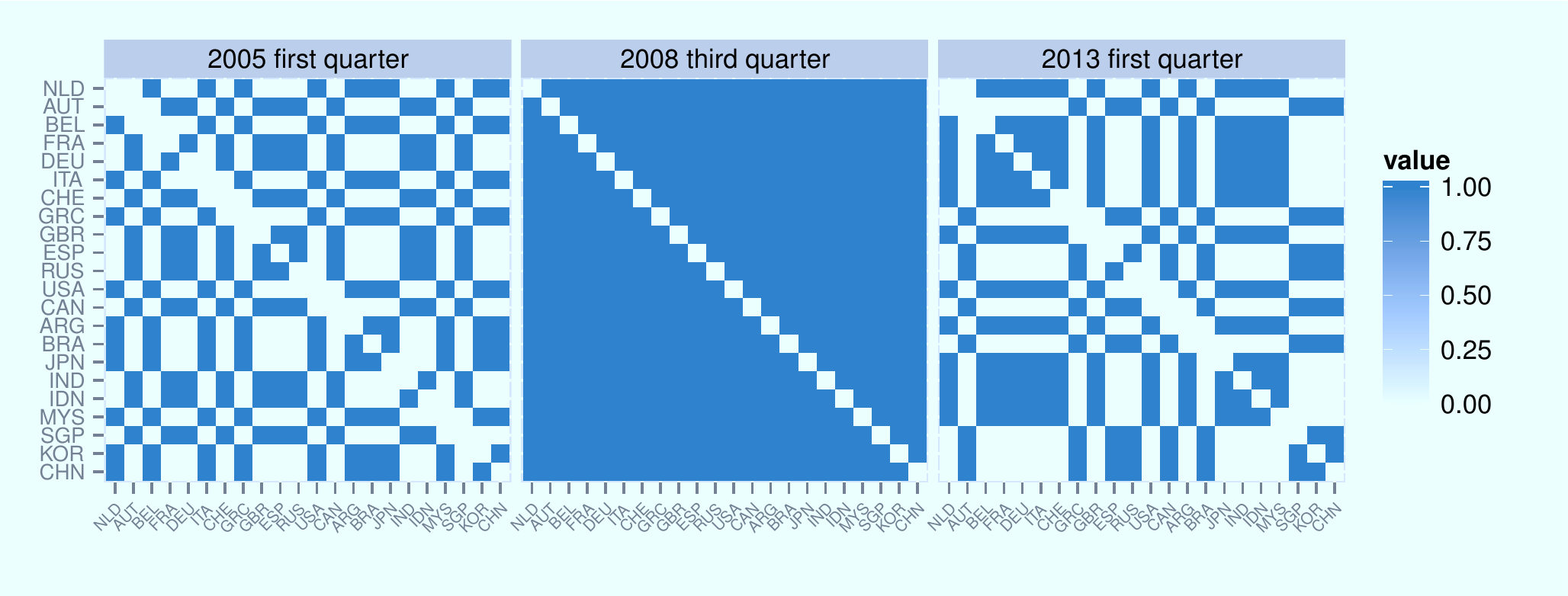}
\caption{{\footnotesize{Response network: dynamic co-movements in world financial markets.}}}
\label{f1}
\end{figure}

\begin{figure}[!h]
\centering
\subfigure{\includegraphics[height=3.7cm, width=10cm]{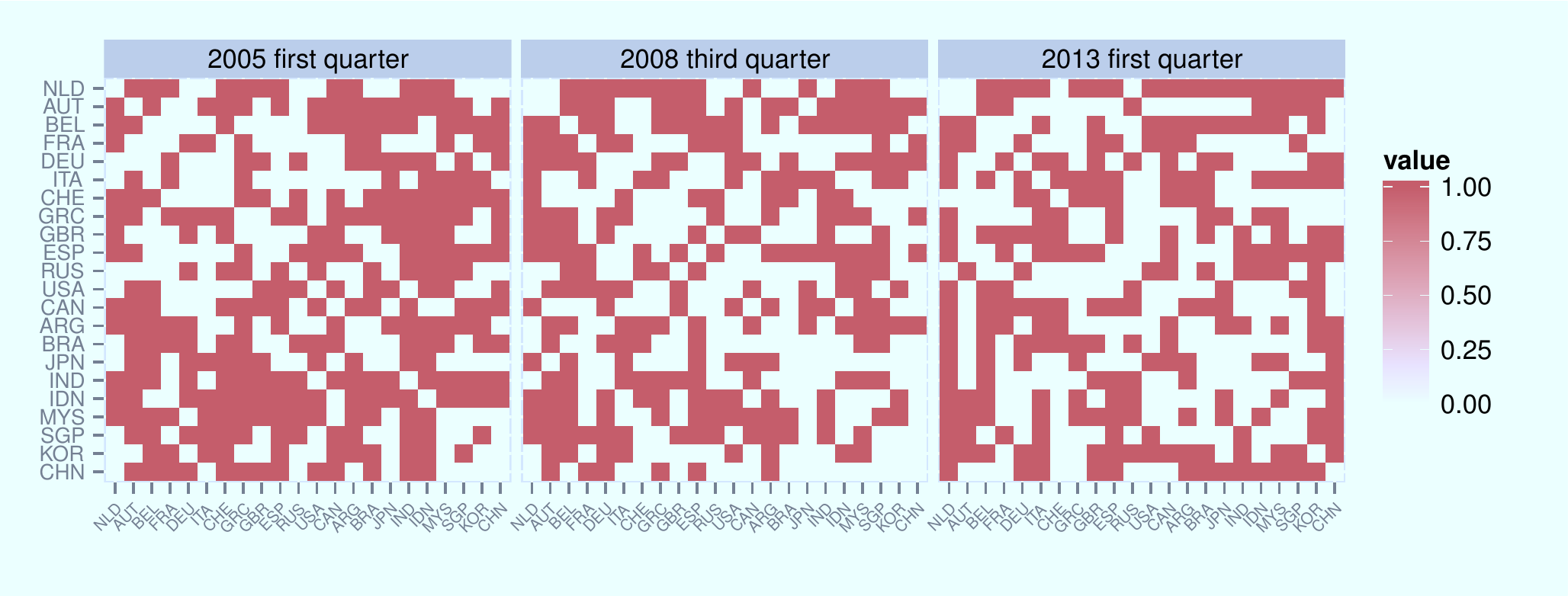}}
\subfigure{\includegraphics[height=3.7cm, width=10cm]{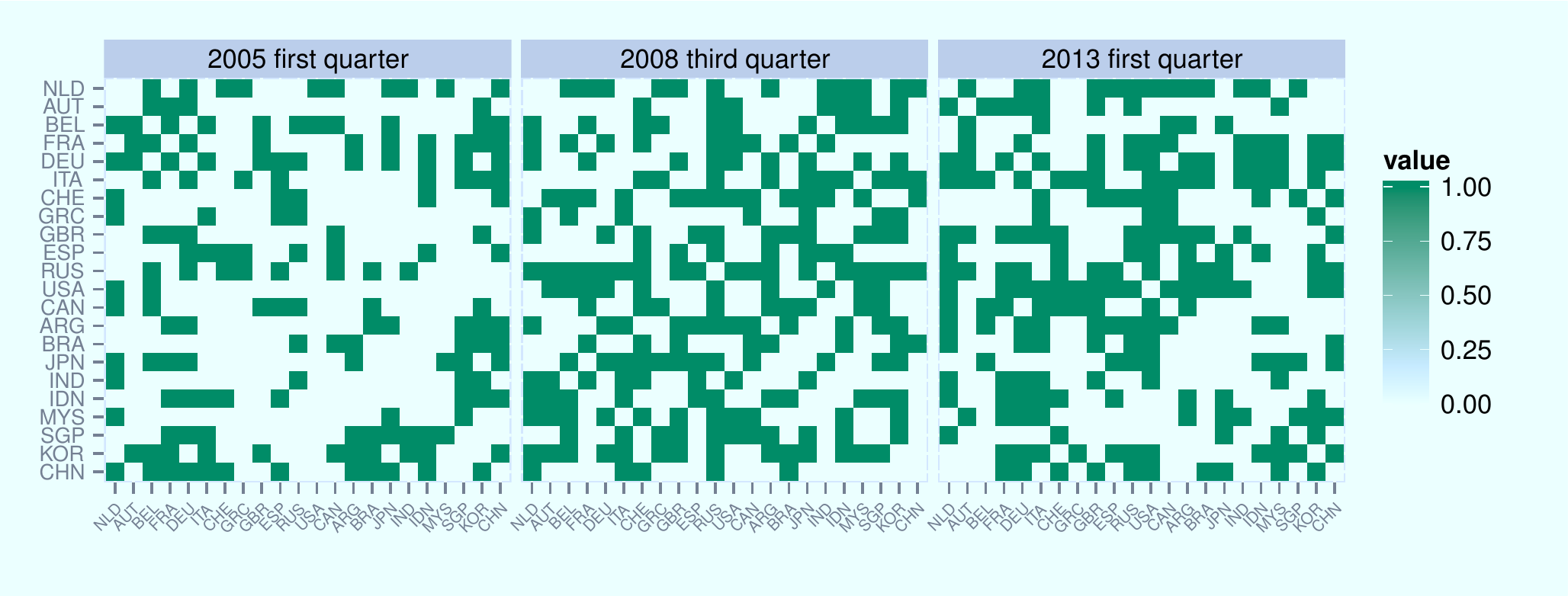}}
\caption{{\footnotesize{Time-varying edge-specific regressors. Upper panels: binary indicators for the presence or absence of a significant increment in material cooperation among pairs of countries. Lower panels: same quantities with respect to verbal cooperation.}}}
\label{f2}
\end{figure}

The increasing interconnection between world financial markets and institutions generated a contagion effect that took shape through the rapid development and spread of the subprime mortgage crisis to the 2008-2012 global recession, which affected the entire world economy and finance. In most cases the recession was manifested through a sharp drop in international trade, low consumer confidence and sovereign-debt crises, which propagated in the subsequent 2010-2012 European sovereign debt crisis affecting mainly Greece, Portugal, Ireland, Spain and Italy, and requiring important material bailout investments by Eurozone. 

Spurred by interest in financial crises and by the need to provide more flexible and accurate statistical analysis of financial systems, a rich variety of statistical methods have been recently developed. Beside descriptive studies interpreting empirical evidences in the light of the key financial events, such as \citet{tay}, there is an abundant literature on model based statistics aimed at exploring co-variations and interconnection structures among financial instruments during recent crises via vector autoregression \citep{long}, vector error correction \citep{ge}, Bayesian factor stochastic volatility models \citep{kas}, locally adaptive factor processes \citep{dur2} and dynamic matrix factorization \citep{sand}. Such applications provide useful overviews on the temporal and geo-economic changes in world financial markets, showing how high volatility phases are directly linked with increasing levels of interdependence. 

Instead of measuring the dynamic dependence structure among a set of financial indices via time-varying covariance or correlation matrices of their corresponding log-returns $R_t=[r_{1,t},\ldots,r_{V,t}]^T$, $ t \in \mathcal{T} \subset \Re^{+}$, we treat co-movements as dynamic relational data and focus inference on the sequence of $V \times V$ time-varying symmetric matrices $\{Y_t, \ t \in \mathcal{T}\}$ having entries $y_{ij,t}=y_{ji,t}=1$ if index $i$ and index $j$ move in the same direction at time $t$, meaning that $r_{i,t}>0$ and $r_{j,t}>0$, or $r_{i,t}<0$ and $r_{j,t}<0$ (indices are similar);  and $y_{ij,t}=y_{ji,t}=0$ if opposite increments are recorded (indices are dissimilar); see Fig.~\ref{f1} for an example. Financial networks provide insight into the factors driving market behavior and risk (see e.g. \citealp{ace} and the references cited therein), but proposals analyzing the impact of other key variables on dynamic variations in financial networks are still lacking. To provide new insights, we exploit the Global Database of Events, Language and Tone to construct two systemic indicators encoding the presence or absence of substantial increments in verbal and material cooperation efforts among pairs of countries as in Fig.~\ref{f2}.  We generalize a recent Bayesian nonparametric model for dynamic relational data to explicitly explore the effects of such covariates on financial interconnections among countries.

\section{Bayesian Dynamic Networks with Time-Varying Predictors}
\subsection{Relevant Contributions}
There is a growing literature on social networks mainly focusing on exponential random graphs  \citep{fra}, stochastic block models \citep{now}, mixed membership stochastic block models \citep{ai} and latent space models \citep{hof}. Current dynamic network models (\citealp{rob}; \citealp{xu}; \citealp{xin}; \citealp{sar}) raise open questions about coherency, flexibility, theoretical properties and computational tractability. Contributions considering edge-specific covariate effects are available in static settings  (see e.g., \citealp{hun}; \citealp{zan}; \citealp{hof}) and developments in the longitudinal framework have been recently explored (\citealp{sni}; \citealp{cra}; \citealp{war}). Such approaches inherit the drawbacks of the dynamic network models they seek to generalize, with only \citet{war} allowing the edge covariate parameters to vary over time via a sequential estimating approach, which does not borrow dynamic information efficiently, and fails to properly propagate uncertainty.  

\subsection{Model Formulation}
We build on the \cite{dur1} nonparametric Bayesian dynamic model for relational data, which efficiently exploits a latent space representation of network data while incorporating time dynamics via Gaussian process (GP) latent factors.  Specifically, the model defines
\begin{eqnarray}
y_{ij,t} | \pi_{ij}(t) &\sim& \mbox{Bern}(\pi_{ij}(t)) \quad t \in  \mathcal{T},
\label{eq1}
\end{eqnarray}
independently for each $i=2,\ldots,V$ and $j=1,\ldots,i-1$, with 
\begin{eqnarray}
\pi_{ij}(t)& = &\frac{1}{1+e^{-s_{ij}(t)}}\enspace,\quad s_{ij}(t)=\mu(t)+x_{i}(t)^T x_{j}(t),\label{eq2}
\end{eqnarray}
where $x_{i}(t)=\left[x_{i1}(t),\ldots, x_{iH}(t)\right]^T$, $i=2,\ldots,V$ and $x_{j}(t)=\left[x_{j1}(t),\ldots,x_{jH}(t) \right]^T$, $j=1,\ldots,i-1$, are the vectors of latent coordinates of unit $i$ and $j$, respectively. 
The link probabilities are estimated via a logistic regression, with $\mu(t)$ a baseline process quantifying the overall propensity to form links in the network across time and $x_i(t)^T x_j(t)$ favoring a higher link probability when units $i$ and $j$ have latent coordinates in the same direction; refer to \cite{dur1} for related theoretical properties.  Bayesian inference proceeds via a data augmentation MCMC algorithm exploiting a representation of \cite{pol}, while allowing estimation of the latent space dimension; see \cite{cho} for recent results on uniform ergodicity
of the resulting algorithm. However, there is no consideration of covariates that may inform about the link probabilities.

We generalize the model to accommodate edge-specific predictors having time-varying coefficients as follows: 
\begin{eqnarray}
s_{ij}(t)&=&\mu(t)+z_{ij,t}^T\beta(t)+x_{i}(t)^Tx_{j}(t),
\label{eq3}
\end{eqnarray}
where $z_{ij,t}=\left[z_{ij1,t},\ldots, z_{ijP,t}\right]^T$ is a $P$-dimensional vector of time-varying edge-specific predictors for units $i$ and $j$ at time $t$ and 
$\beta(t)=\left[\beta_1(t),\ldots, \beta_{P}(t)\right]^T$ are the corresponding dynamic coefficients. This allows the proximity between units $i$ and $j$ at time $t$ to depend on predictors in a manner that varies smoothly with time.  These time-varying coefficients are of substantial inferential interest.

In our motivating finance application, latent coordinates may represent unexpected inflation and investors expectations, respectively, favoring indices of countries with features in the same directions to co-move, and countries with opposite unexpected inflation and investors expectations to move in different directions with higher probability. Additionally, we allow the presence of a significant increment in verbal or material cooperation relations among pairs of countries to further increase or decrease the co-movement probability proportionally to its corresponding time-varying coefficient. 

\subsection{Prior Specification and Posterior Computation}
We follow \cite{dur1} in assuming Gaussian process priors for the baseline process $\mu(\cdot)$ and the time-varying latent features $x_{ih}(\cdot)$, $i=1,\ldots,V$ and $h=1,\ldots,H$, with the priors for the feature curves designed to adaptively shrink unnecessary latent curves to zero, allowing inference on the dimension of the latent space.
To flexibly model the effect of the edge-specific dynamic regressors, we consider independent Gaussian process priors
\begin{eqnarray}
\beta_p(\cdot)\sim \mbox{GP}(0,c_{p}) \quad p=1,\ldots,P,
\label{eq4}
\end{eqnarray}
with $c_{p}$ a squared exponential correlation function  $c_{p}(t,t')=\exp(-\kappa_{p} ||t-t' ||^{2}_{2})$, which allows for continuous time analysis and unequal spacing.

Posterior computation is performed generalizing the \cite{dur1}  Gibbs sampler to update the vector of time-varying regression coefficients. This is accomplished by modifying the steps outlined in \cite{dur1} to account for the new definition of $s_{ij}(t)$ as in (\ref{eq3}) and by considering a further step updating the dynamic regression coefficients $\beta_{p}(t)$, $p=1,\ldots,P$, exploiting the conjugacy provided by P\'olya-gamma data augmentation and the GP assumption. 

Specifically, letting $\mathcal{T}=\{ t_1, \ldots, t_N\}$ denote the time grid on which networks are observed, sample $\beta_p=[\beta_p(t_1),\ldots,\beta_p(t_N)]^T$, for each $p=1, \ldots, P$, from its full conditional $N$-variate Gaussian posterior

\begin{eqnarray*}
 \beta_p \mid -   \sim \mbox{N}_{N}\left(\Sigma_{\beta_p}\left[ \begin{array}{c}
\sum_{i=2}^{V}\sum_{j=1}^{i-1}z_{ijp,t_1}\left(y_{ij,t_1}-1/2-\omega_{ij,t_1}\nu_{ijp,t_1}\right)\\
\vdots\\
\sum_{i=2}^{V}\sum_{j=1}^{i-1}z_{ijp,t_N}\left(y_{ij,t_N}-1/2-\omega_{ij,t_N}\nu_{ijp,t_N}\right)
 \end{array} \right],\Sigma_{\beta_p}\right),
\end{eqnarray*}
where $\nu_{ijp,t}=\mu(t)+z_{ij(-p),t}^T\beta_{(-p)}(t)+x_{i}(t)^Tx_{j}(t)$, $t=t_1, \ldots, t_N$, with $z_{ij(-p),t}$ and $\beta_{(-p)}(t)$, the edge-specific predictor vector and the corresponding dynamic coefficients, respectively, with the $p$-th element held out. In addition the posterior covariance matrix is $\Sigma_{\beta_p}=\left\{\mbox{diag}\left(\sum_{i=2}^{V}\sum_{j=1}^{i-1}z_{ijp,t_1}^2\omega_{ij,t_1},\dots,\sum_{i=2}^{V}\sum_{j=1}^{i-1}z_{ijp,t_N}^2\omega_{ij,t_N} \right)+ K_{p}^{-1}\right\}^{-1}$, with $K_{p}$ the Gaussian process covariance matrix with entries $(K_{p})_{ij}=\exp(-\kappa_{p} ||t_i-t_j ||^{2}_{2})$.

\section{Co-movements Regression Financial Network}
\begin{figure}[t]
\centering
\includegraphics[height=9cm, width=10cm]{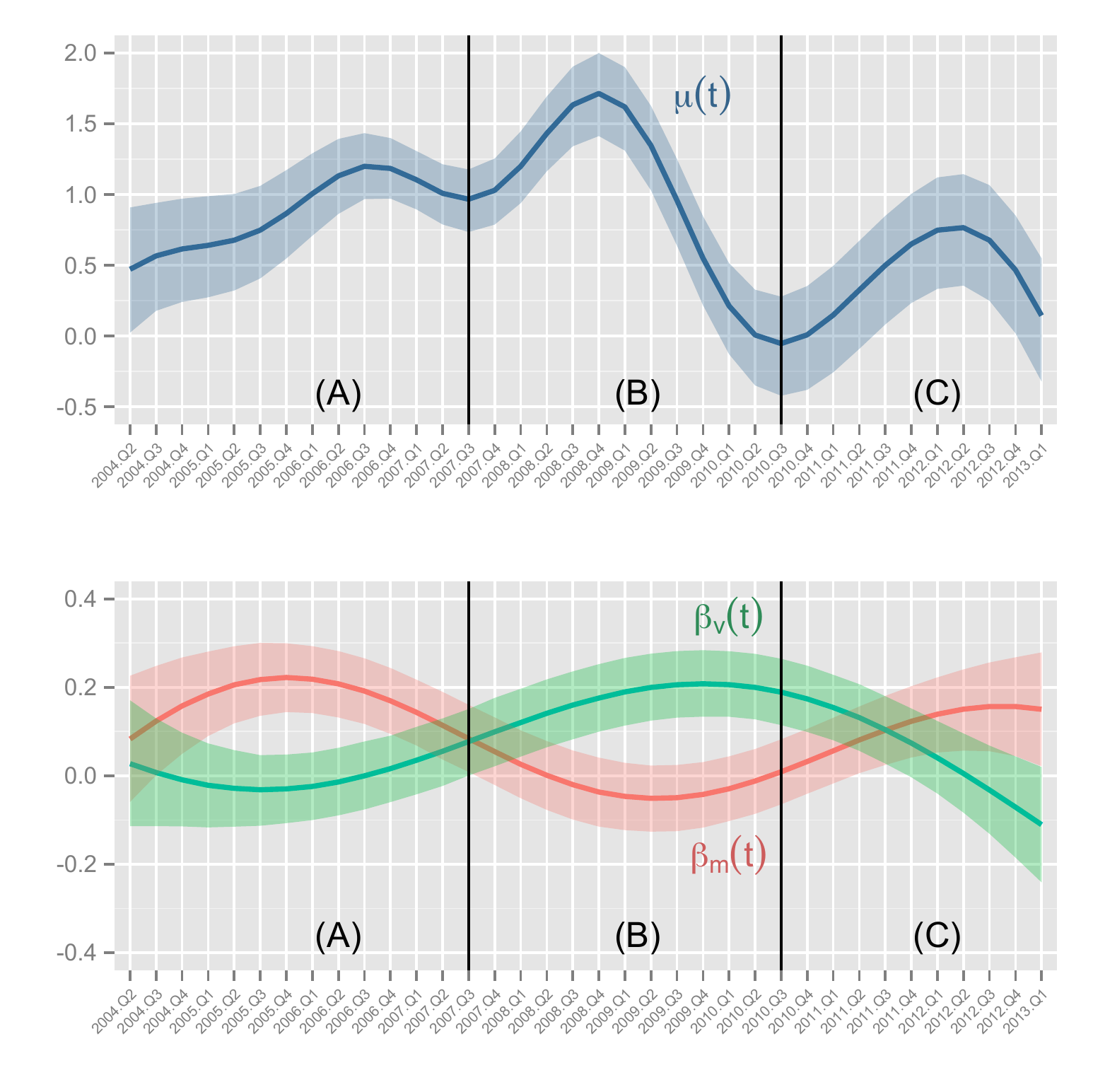}
\caption{{\footnotesize{Upper plot: point-wise posterior mean for the baseline $\mu(t)$ (colored blue line), and 0.95 highest posterior density (hpd) intervals (colored blue areas). Lower plot: same quantities for the time varying verbal (green) and material (red) cooperation regression parameters. (A) Growth and burst of the United States housing bubble, (B) Global financial crisis, (C) Greek debt crisis and worsening of the  European debt crisis.}}}
\label{f3}
\end{figure}

\subsection{Data Processing}
We combine Yahoo Finance data and the historical event dataset from the  Global Database of Events, Language and Tone, to dynamically model the financial network among the main $V=22$ national stock market indices and learn whether and to what extent substantial increments in verbal and material cooperation relations affect the probability to co-move.  Specifically, we construct the response co-movement network $\{Y_{t}, t =1,\ldots,36 \}$ using quarterly log-returns from the second quarter of 2004 to the first quarter of 2013, available at {\url{http://finance.yahoo.com/}}, with data at the last time held out to assess predictive performance.  For dynamic regression we consider two binary edge-specific variables, $z_{ijv,t}$ and $z_{ijm,t}$, indicating the presence or absence of a substantial increment in verbal or material relations, respectively, between country $i$ and $j$ at time $t$. To construct $z_{ijv,t}$ (the same procedure holds for $z_{ijm,t}$), we exploit the indicator \texttt{IsRootEvent} to restrict our analysis to the subset of important relations in the historical event datasets available at {\url{http://www.gdeltproject.org/data.html}} and use variable \texttt{QuadClass} to compute the difference between the total number of verbal cooperation and verbal conflict events recorded among countries $i$ and $j$ at time $t$, for each pair of countries and quarter $t$. To mitigate the explosive non-stationary trend of such quantities and the very different scale of each time series, we use standardized increments obtained by computing the first difference of each time series and standardizing the latter with its unconditional standard deviation. Finally, we define as `substantial' an increment greater than the average of all standardized first differences at time $t$, assigning $z_{ijv,t}=1$ in such cases and $z_{ijv,t}=0$, otherwise, for  $i=2,\ldots,22$ and $j=1,\ldots,i-1$ and $t =1,\ldots,36$. This procedure allows us to carefully define our dynamic edge-specific binary indicators encoding verbal and material relative proximity measures among countries.

\subsection{Model Interpretation}

We apply model (\ref{eq1}), with latent similarity measures given by
\begin{eqnarray}
s_{ij}(t)&=&\mu(t)+z_{ijm,t}\beta_{m}(t)+z_{ijv,t}\beta_{v}(t)+x_{i}(t)^T x_{j}(t)\enspace,
\label{eq5}
\end{eqnarray}
with $i=2,\ldots,22$, $j=1,\ldots,i-1$ and $t=1,\ldots,36$. For inference we set $H=15$, length scales $\kappa_{\mu}=0.02$, $\kappa_{x}=0.01$, $\kappa_{\beta_{m}}=\kappa_{\beta_{v}}=0.01$ and fix $a_1=a_2=2$ for the shrinkage parameters. We consider $5{,}000$ Gibbs iterations, with a burn-in of $1{,}000$. Mixing was assessed via estimating effective sample sizes for the quantities of interest, represented by $\pi_{ij}(t)$, $\mu(t)$, $\beta_{m}(t)$ and $\beta_{v}(t)$ for $i=2,\ldots,22$, $j=1,\ldots,i-1$ and $t=1,\ldots,36$ after burn-in. Most of these values were around $1{,}500$ out of $4{,}000$, suggesting good mixing. We interestingly find that the first two latent factors are the most informative, with the remaining redundant latent features consistently shrunk to zero. A similar insight was highlighted by \cite{fam}, showing three common latent factors underlying stock returns.

\begin{figure}
\centering
\includegraphics[height=4.5cm, width=12.5cm]{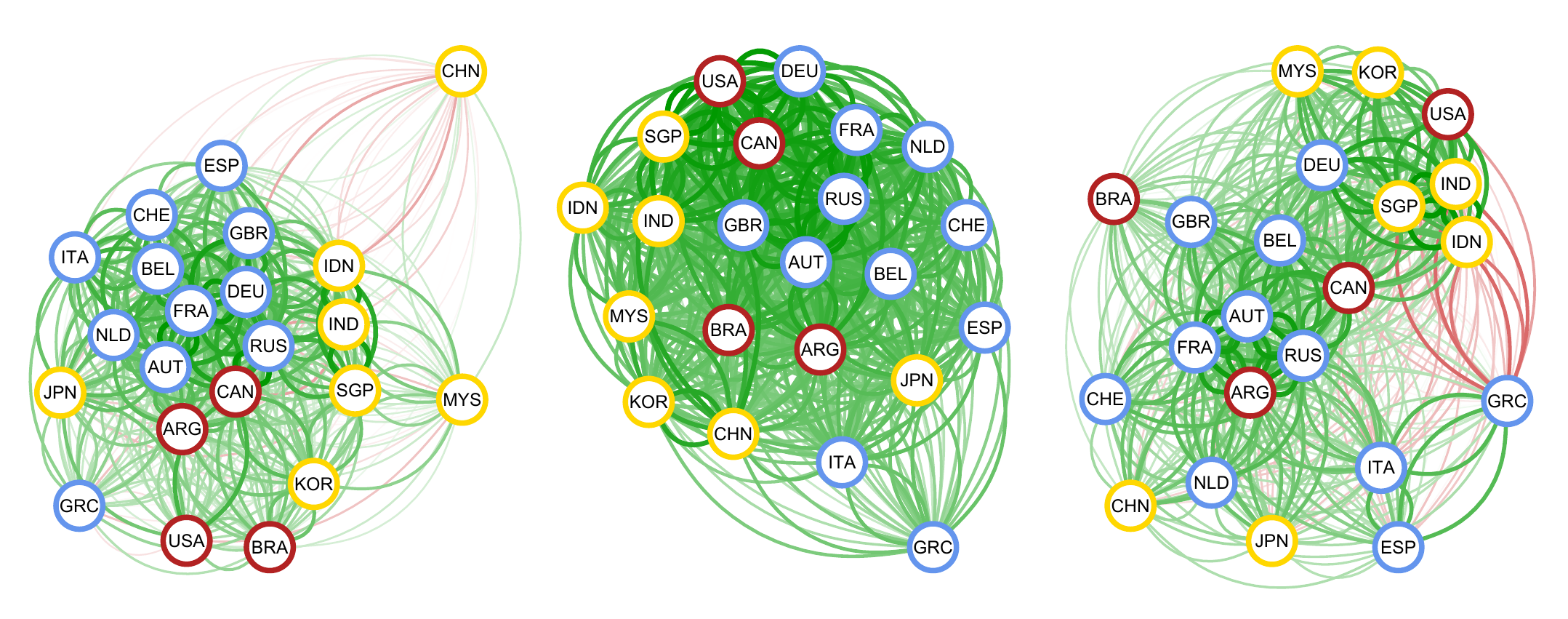}
\caption{{\footnotesize{Weighted network visualization with weights obtained averaging $\hat{\pi}(t)$ over all the time window (left),  over the period of the 2008 global financial crisis (middle) and over the period of Greek debts crisis (right).  Edge dimensions are proportional to the corresponding value of the averaged co-movement probability, with colors going from red to green as the corresponding weight goes from $0$ to $1$. Blue, red and yellow nodes represent European, American and Asian countries.}}}
\label{f5}
\end{figure}

The estimated peaks in overall co-movement propensity $\mu(t)$ in Fig.~\ref{f3}, during the most acute phases of the crisis and the change of regime at the burst of the United States housing bubble (A), provide empirical support of contagion effects in correspondence to the global financial crisis (B) and the European sovereign-debt crisis (C). Figure \ref{f5} shows three weighted networks, with weights given by the estimated co-movement probability averaged over different time windows. As expected geo-economic proximity among countries is manifested through tighter networks, with  Japan most closely related to western economies, China having lower interconnections with other economies and stronger networks for European markets and Asian Tigers, respectively. The denseness of the network during the 2008 global financial crisis provides a further evidence in support of international financial contagion effects; while the low connections among Greece and almost all the other financial markets during the Greek debt crisis, and the strong relations with Spain and Italy representing the countries most affected by the financial instability during such phase, provide interesting insights on the features of the 2010-2012 European sovereign-debt crisis.

These results confirm \cite{dur1} findings, and agree with theories on financial crises \citep{cla} and recent applications in \cite{kas}, \cite{dur2}. However, the novelty and significance is in providing a quantitative overview on the dynamic effects of substantial increments in verbal and material relations on co-movement propensity. The generalization of \cite{dur1} to include information provided by edge-specific time-varying regressors allow us to improve predictions with an area underneath the ROC curve of $0.85$ for data at the last time, and to provide additional important insights as highlighted in 
 the last plot of Fig~\ref{f3}. 

Coefficients evolve in general on positive values inducing higher probability to co-move for countries with substantial increments in verbal or material relations.  
Trajectories cross at the beginning and end of the most severe phase of the 2008 global financial crisis. Such results highlight the importance of time-varying coefficients, illustrating a 
significantly higher impact of increments in material relations on co-movement probabilities during the growth and burst of the United States housing bubble and an opposite behavior when the time window of the global financial crisis is considered.   Recalling the key events during the period considered, higher values of the material coefficient before 2008 are in line with the ``originate and distribute'' banking model and predatory lending during the United States housing bubble, which favored the creation of securities of great complexity (such as RMBS and CDO), and stimulated large capital exchanges inflating the network of material relations among financial institutions worldwide and increasing its impact on the behavior of financial markets. Consistently with this interpretation, the material coefficient peaks in proximity of the bubble burst and is closer to zero during the most acute phase of the global crisis when the proliferation of meetings and global conferences between governments and financial institutions, and the lack of material funds to invest in foreign markets,
lead to an increase of the verbal coefficient.  The difference among the two coefficients is less evident in the time-window of the European sovereign debt crisis, which is characterized by important material bailout investments by the Eurozone institutions, such as European Financial Stabilization Mechanism, the European Financial Stability Facility, and the International Monetary Fund.

\section{Conclusion}
Developing statistical models to flexibly learn time-varying network structures, while inferring the effects of additional variables, is a key issue in many applied domains.  It is increasingly common to have data available on dynamic networks and related node features.  In addition, viewing data through a network lens can add substantial new insights, as we have illustrated in our motivating finance application.  It is interesting to consider further generalizations and modification of our nonparametric Bayes dynamic latent space model.  For example, issues may arise in using a smooth Gaussian process, particularly as the number of time points at which observations are collected increase.  Firstly, there are computational issues that arise due to bottlenecks and ill-conditioning problems in matrix inversion.  These issues can be directly addressed using low-rank approximations to the GP, such as the random projection method of \cite{ban}.  In addition, we have relied on GPs with stationary covariance functions, implying time-constant smoothness in the latent processes and coefficient curves.  In finance and other network applications, such as neurosciences, we expect smoothness to vary over time.  This motivates the use of time-varying smoothness, potentially accomplished via stochastic differential equation models, such as the nested Gaussian process of \citet{zhu}.  Finally, it is important to develop scalable computational methods that can accommodate very large networks having huge numbers of nodes; in this respect, models and computational approaches that exploit sparsity provide a promising direction.

\bibliography{biblio}

\end{document}